\documentclass{article}
\usepackage{graphicx}
\usepackage{amsmath,amsthm}
\usepackage{natbib}
\usepackage{rotating}
\usepackage{setspace}
\usepackage{fullpage}

\newcommand{\be}{\begin{equation}}
\newcommand{\ee}{\end{equation}}

\newcommand{\btheta}{\mbox{\boldmath{$\theta$}}}
\newcommand{\bdelta}{\mbox{\boldmath{$\delta$}}}

\newtheorem{exmp}{Example}
\newtheorem{res}{Result}
\newtheorem{defin}{Definition}
\bibpunct{(}{)}{;}{a}{}{,}
\newcommand{\ssection}[1]{\section[#1]{\centering #1}}

\begin{document}
\bibliographystyle{abbrvnat}

\title{Limiting the Shrinkage for the Exceptional by Objective Robust Bayesian Analysis: the \emph{``Clemente Problem"} }
\author{Luis R. Pericchi and Mar\'{\i}a-Egl\'ee P\'erez \footnote{Luis R. Pericchi is Professor, Department of Mathematics, University of Puerto Rico, Rio Piedras Campus, PR 00931 (e-mail: luarpr@gmail.com). Mar\'{\i}a-Egl\'ee P\'erez is Assistant Professor, Department of Mathematics, University of Puerto Rico, Rio Piedras Campus, PR 00931 (e-mail: meglee@uprrp.edu). This work was sponsored in part by NIH Grant: P20-RR016470. M.E. P\'erez research was also sponsored by NSF Grant: HRD 0734826. The authors want to dedicate this work to Roberto Clemente Walker, first latin-american to reach the Baseball Hall of Fame.}}
\date{September 2011}
 \maketitle

%
%
%
%

\doublespacing

\begin{abstract}
Modern Statistics is made of the sensible combination of direct evidence
(the data directly relevant or the ``individual data") and indirect
evidence (the data and knowledge indirectly relevant or the ``group
data"). The admissible procedures are a combination of the two sources of
information, and the advance of technology is making indirect evidence
more substantial and ubiquitous. It has been pointed out however, that in
``borrowing strength" an important problem of Statistics is to treat in a
fundamentally different way exceptional cases, cases that do not adapt to
the central ``aurea mediocritas". This is what has been recently coined as
``the Clemente problem" in honor of R. Clemente, an exceptional batter
\citep{Efr09}. In this article we put forward that
the problem is caused by the simultaneous use of square loss function and
conjugate (light tailed) priors which is the usual procedure.  We propose
in their place to use robust penalties, in the form of losses that
penalize more severely huge errors, or (equivalently) priors of heavy
tails which make more probable the exceptional. Using heavy tailed prior
we can reproduce in a Bayesian way,  Efron and Morris' ``limited translated
estimators" (with Double Exponential Priors) and ``discarding priors
estimators" (with Cauchy-like priors) which discard the prior in the
presence of conflict. Both Empirical Bayes and Full Bayes approaches are
able to alleviate the Clemente Problem and furthermore beat the
James-Stein estimator in terms of smaller square errors, for sensible
Robust Bayes priors. We model in parallel Empirical Bayes and Fully
Bayesian hierarchical models, illustrating that the differences among
sensible versions of both are relatively small, as compared with the effect due to
the robust assumptions. We propose a heavy tailed (scaled) Beta2 distribution
for (squared) scales that arises naturally as an alternative to the
usual Inverted-Gamma distribution. The combination of a Cauchy Prior
for location and Beta2 for square scales, yields  novel closed form
priors for location, extremely suitable for Objective Robust
Bayesian Analysis (ORBA).

\end{abstract}

\ssection{SHRINKAGE ESTIMATORS THAT BORROW STRENGTH FROM INDIRECT EVIDENCE: TOO MUCH OF A GOOD THING?}
\subsection{An Historical Example}
\citet{EM75} obtained a sample of batting averages for 18 baseball player
during the 1970 season. They used the average obtained during the first 45
at-bats for predicting the batting
average for the rest of the season for each player.\\
In Table \ref{tab:data}, the relevant data are presented. Direct evidence
is the observed individual average, thus the temptation to predict by the
observed individual average, although it is known that this estimator is
inadmissable. This is a very bad practical predictor in this case, and
this has been corroborated in other cases, see \citet{Br08}. At the other
end, the indirect evidence
comes in the form of the overall mean $M=0.265$ of all batters. \\
This ``pure indirect evidence" estimator is surprisingly good in this
case, and far better than the ``pure direct evidence", MLE or naive
estimator as Brown calls it. However, both intuition and theory point to a
sensible combination of the two sources of evidence to improve overall
predictions. The problem is then: How much to weight direct and indirect
evidence in each individual case? Wouldn't it be reasonable to weight less
the common indirect evidence, when there is reason to believe that the
individual is exceptional? This was the original motivation of Efron and
Morris for their clever (and ``{\it{ad-hoc}}") ``limited translation
estimators" \citep{EM72}.

\subsection{Efron and Morris set up}
The initial assumption about the data in Efron and Morris (1972) is:

\[ Y_i \sim \frac{1}{45}\mbox{Bin}(45,p_i)\]

\noindent where $Y_i$ is the batting average for the first 45 at-bats, and
$p_i$ depends on each player's ability.

The batting average for the rest of the season, $R_i$ can be
modeled as

\[ R_i \sim \frac{1}{n_i} \mbox{Bin}(n_i,p_i) \]

\noindent where $n_i$ is the number of at bats for player $i$ during the
remainder of the season.

They applied a variance stabilizing transformation to $Y_i$,

\[ X_i = \sqrt{45}\arcsin{(2Y_i-1)} \]

\noindent so that $X_i \sim N(\mu_i,1)$, with $\mu_i$ approximately equal to the transformed value of $p_i$. The interest is in predicting the final individual batting average for the rest of the season. In the sequel, we will use this transformed variable.

The analysis of these baseball data by Efron
and Morris (1972) was widely cited and remains one of the
clearest expositions in favor of combining { \em
``indirect evidence"}, with { \em ``direct evidence"}, a practice often
termed ``borrowing strength" and ``shrinkage estimation".

\subsection{The Clemente Problem}
In his talk at the '09 Objective Bayes Conference June 2009,
{\it{``The Future of Indirect Evidence"}} Professor Bradley Efron exposed
a fundamental problem: {\it{``The Clemente Problem: How to protect
atypical cases from too much indirect evidence ?"}}. Professor Efron is
referring to the Puerto Rican sportsman Roberto Clemente, an outstanding
batter and human being, who had the highest batting average of the list of
18 players. After the first $45$ turns, Clemente had a score of $0.400$,
or $40$\% of hits. Even though shrinking to a general mean improves the
overall prediction of the 18 batters, for Clemente his score was predicted
as $0.290$. The atypical Clemente was not protected from ``{\it{too much
of a good thing}}", and his personal prediction was very poor: he finished
with a score of $0.346$, much higher than the predicted. The problem lies
in the fact that the usual method shrinks a fixed proportion to all
players, see equation (\ref{eq:fixprop}) below. It does not make any
exception, for the too good or too bad batters. This is a logical
consequence of the assumptions made, since it corresponds to an optimal
decision in Decision Theory. So, in {\it{``What If?"}} mode of thought if
the logical consequences are not {\it{``pleasant to the mind"}}, {\it{a
fortiori}} assumptions have to be changed. Fixed proportion estimators are
not robust in the sense that the amount of shrinkage is not limited, that
is the potential influence of indirect evidence is unbounded, or using a
metaphoric expression the procedure is ``{\it{myopic}}" to the conflict
between the bulk of the data and the individual.

\subsection{Robust Penalties}
Our starting point is re-analyzing the Loss (or minus Utility)
function. In decision analysis the Square Loss is by far the most
used (or over-used?) and the Clemente problem is (in part) a
coherent consequence of its assumption. To see this, we recall the
following result in Decision Theory.
\begin{res}
Suppose a function of the parameter $\btheta$,  $g(\btheta)$, is being
estimated by $\bdelta(X_1,\ldots, X_m)$. Assume the weighted square loss function: \[
L(g(\btheta),\bdelta)=\sum_{i=1}^m L_i(g(\theta_i),\delta_i({\bf X})) = \sum_{i=1}^m w(\theta_i)\cdot
(\delta_i({\bf X})-g(\theta_i))^2), \,\, w(\theta_i) \geq 0. \] Then the optimal Bayes
estimator is: \be \delta_i({\bf X})=\frac{E[w(\theta_i)\cdot
g(\theta_i)|\mbox{data}]}{E[w(\theta_i)|\mbox{data}]}. \label{eq:optim} \ee
\end{res}
\noindent {\bf Proof:} \citet[p.~47]{Fer67}.

 Although this result may be termed as classical, its
statistical consequences have not being fully appreciated. In fact
(\ref{eq:optim}) invites two strategies: the first is to weight the
square loss and use (\ref{eq:optim}) as the individual estimator and the
second is to change the prior in a way suggested by (\ref{eq:optim}), and
keep the square loss function. These are two different view points that
shed different lights and possibilities. In this paper we highlight
assumptions to get robust solutions both ways.

\ssection{HEAVIER THAN QUADRATIC LOSSES}
For simplicity we will assume we are estimating $\btheta$,  that is,
$g(\btheta)=\btheta$.

To diminish the shrinkage on the extremes, the loss function (centered around the overall group location) has to
penalize errors far from the overall location more heavily than square loss. We also need an
origin $M$ where to anchor the weighting function $w(\theta_i)$. This origin $M$ might be chosen subjectively, using prior knowledge and experience or empirically, and could be interpreted as the mean of the $\theta_i$'s.


\begin{exmp}[Exponentially weighted loss:]

 \[ L_i^{exp}(\theta_i,\delta_i({\bf X}))=\exp[r\cdot|\theta_i-M|]
\cdot (\theta_i-\delta_i({\bf X}))^2, i=1,\ldots, m, r>0 \]

\end{exmp}
Calculation yields as the optimal estimator in (\ref{eq:optim}) \[
\delta_i^{exp}({\bf X})=\frac{a[b(\mu_{1i}-rv)-c]+a'[c'+(1-b')(\mu_{1i}+rv)]}{ab+a'(1-b')},
\] where $a=\exp(r(M-\mu_{1i})+vr^2/2))$,
$b=\Phi(\frac{(M-(\mu_{1i}-rv)}{\sqrt{v}})$,
$c=\sqrt{v}\phi(\frac{M-(\mu_{1i}-rv)}{\sqrt{v}})$,
$a'=\exp(r(\mu_{1i}-\tilde{\theta})+vr^2/2))$,$b'=\Phi(\frac{\tilde{\theta}-(\mu_{1i}+vr)}{\sqrt{v}})$,$c'=\sqrt{v}\phi(\frac{\tilde{\theta}-(\mu_{1i}+rv)}{\sqrt{v}})$,
and $\mu_{1i}, v$ are the posterior mean and variance respectively of a
Normal likelihood with a Normal prior.

This and other loss functions can lead to tractable results, but it is
more convenient for the purposes of the present paper to work with losses
that have also a direct interpretation in terms of heavy tailed (robust)
priors, which penalizes more heavily discrepancies and that are naturally
scaled.

\begin{exmp}[Cauchy over Gaussian Loss:] Here we place a Cauchy
density over a Gaussian both centered at $M$ with matched interquartile
range. \[
L_i^{CG}(\theta_i,\delta_i({\bf X}))=\frac{\mbox{Cauchy}(\theta_i|M,1)}{N(\theta_i|M,
2.19)} \cdot (\theta_i-\delta_i({\bf X}))^2. \]
\end{exmp}

Figure 1 shows the Cauchy over Gaussian loss, being quite close to square
loss around zero but growing fast without bound for ``exceptional" values.
\begin{res}
In fact the optimal estimator (\ref{eq:optim}) under the
$L_i^{CG}(\theta_i,\delta_i({\bf X}))$ is the posterior expectation under a Cauchy prior,
since, using the Gaussian with mean $M$ and variance $2.19$ as a prior

 \begin{eqnarray*}
 \delta({\bf X})=\frac{E[w(\theta_i)\cdot \theta_i|{\bf X}]}{E[w(\theta_i)|{\bf X}]}&=&
 \frac{\int \frac{\mbox{Cauchy}(\theta_i|M,1)}{N(\theta_i|M, 2.19)} \theta_i \cdot \pi(\theta_i|{\bf X})}{\int \frac{\mbox{Cauchy}(\theta_i|M,1)}{N(\theta_i|M, 2.19)} \pi(\theta_i|data)}\\
 &=& E[\theta_i|\mbox{{\bf X}, Cauchy Prior}]
 \end{eqnarray*}

 \end{res}

 In words, the optimal estimator under the Robust Cauchy over Gaussian penalty
(and Gaussian prior) is the posterior expectation under a Cauchy Prior,
since with square loss the optimal estimator is the posterior expectation.

So a bridge has been established between Robust Losses and Robust Priors
through equation (\ref{eq:optim}). We call it comprehensive  Robustness,
the use of Robust Loss Functions or the use of Robust (heavy tailed)
Priors. For convenience, now we go to the ``Robust Prior Route".

In what follows we try different models, in two methods: Empirical Bayes
and Fully Robust Bayes, trying to solve, or at least alleviate the
Clemente problem, that is, the lack of robustness of exponential family
models with conjugate (light tailed) priors and squared loss function.

\ssection{ROBUST CONSEQUENCES OF ROBUST (HEAVILY TAILED) PRIORS}

In this section we motivate briefly the use of heavy tailed priors as
a tool for robustness. Back in the situation of subsection 1.2,  let us make the usual assumption of a Normal
Likelihood and Prior:
\begin{eqnarray}
X_i & \sim & \mbox{Normal}(\mu_i,1), \;\; i=1, \ldots , 18 \label{eq:lik}\\
\mu_i & \sim & \mbox{Normal}(\mbox{M},\sigma_0^2),\label{eq:normprior}
\end{eqnarray}
where $(\mbox{M},\sigma_0^2)$ has been assigned, for instance via an
Empirical Bayesian method \citep[see][]{EM72} or using previous year batting averages (for 1969, the global batting average was $0.248$, see for example \\ {\tt http://www.baseball-reference.com/leagues/MLB/1969.shtml}). These assumptions coupled with square error loss, lead to the posterior conditional expectation as the optimal
estimator, which can be written as: \be E(\mu_i|x_i)=x_i +
\frac{1}{1+\sigma^2_0} (M-x_i). \label{eq:fixprop}\ee It is convenient here to define shrinkage as  $|E(\mu_i|x_i)-x_i|$.There are
several ways to analyze the lack of robustness of (\ref{eq:fixprop}), but
the one that is more relevant here: {\it{all batters either exceptional or
average are shifted to the mean of the means $M$ a {\bf{fixed}} }} proportion
 $c=1/(1+\sigma^2_0)$, so the shrinkage is
$c\cdot |M-x_i|$. That is the ``Clemente Problem". Notice that
increasing the prior variance $\sigma^2_0$ is {\bf{not}} a fix to the
problem: it would reduce the shrinkage certainly, but to all batters in
equal proportion, even to those who are not exceptional. We are not proposing a indiscriminated reduction of the shrinkage, but rather a differential shrinkage. The posterior
mean (\ref{eq:fixprop}) is {\it{myopic}} to the exception. Now
(\ref{eq:fixprop}) is the logical consequence of the assumptions. Thus the
only logical way to change it is to change the assumptions. We could
change the loss, but equivalently we change the tail behavior of the
priors: Using flatter tails, is giving to Bayes Theorem the input that the
exceptional is more likely.

Our first alternative is a Double Exponential prior:
\begin{eqnarray}
\mu & \sim & \mbox{DE}(\mbox{M},\nu_0)=\frac{1}{\nu_0 \sqrt{2}}
\exp(-\frac{\sqrt(2)}{\nu_0} |\mu-\mbox{M}|),\label{eq:deprior}
\end{eqnarray}
where $\nu_0=\frac{\sqrt(2) \sigma_0}{\log(2)} \Phi^{-1}(0.75)$, to
match the quartiles of the Normal prior. Finally for even heavier
tails we explore with a Cauchy prior.
\begin{eqnarray}
\mu & \sim & \mbox{C}(\mbox{M},\gamma_0)=\frac{1}{\pi
\gamma_0}\frac{1}{1+(\mu-\mbox{M})^2/\gamma_0^2},\label{eq:cauprior}
\end{eqnarray}
where $\gamma_0=\sigma_0 \Phi^{-1}(0.75)$, again to match Normal
quartiles.

For exact and approximate results with these, and other priors, see for
example \citet{PS92}, but Figures 2 and 3 tell the story. In Figure 2 the
data are kept fixed at zero and the prior location is moved as to create a
conflict between one data point and a prior. With a Normal prior the
shrinkage grows linearly without bound. The other two priors yield robust
estimators. For a Double Exponential prior, the posterior expectation
become essentially a ''{\it{limited translation estimator}}" in Efron and
Morris' terminology. The influence of the overall mean $\mbox{M}$ is
bounded and monotonic. Finally for the Cauchy prior, the posterior
expectation is {\bf{not}} monotonic in the conflict between the MLE and
the General Mean. Furthermore, the prior is progressively discarded in
favor of the MLE, as the general mean and MLE diverge. It is also quite
interesting that the shrinkage of the two robust priors is almost the same close to the center, that is around the general average $M$, and actually they give more shrinkage near
the center than the Normal.

In Figure 3 we are showing the actual values of the transformed data
$X_i, \;\; i=1,\ldots,18$ in the x axis. The prior location $\mbox{M}$
is fixed at the sample overall average. Around the middle-ground the three
estimators are very close together. But at some point, the robust
estimators separate from the Normal, in the direction of the MLE, being
the adjustment towards the MLE of the Cauchy somewhat stronger than that
of the Double-Exponential.

\ssection{MODELS}
We will consider two types of models, Empirical Bayes and Full Bayesian
strategies.

\subsection{Empirical Bayes Strategy}

Following \citet{EM75}, general location and scale parameters are
calculated from the sample: $ \mbox{M} = \bar{X} =-3.3166$ and
$\tilde{\sigma}^2$ such that $ \frac{1}{(1+\tilde{\sigma}^2)} =
\frac{k-3}{\sum_{i=1}^k(X_i - \bar{X})^2}$, so $\tau = (\sigma^2)^{-1}=
3.7853$.

{\bf{Models 1, 2 and 3}}, are defined with the likelihood (\ref{eq:lik})
and respectively priors (\ref{eq:normprior}), (\ref{eq:deprior}) and
(\ref{eq:cauprior}). The justification is as follows: the first prior is
the original analysis by Efron and Morris, Double Exponential and Cauchy
priors are two heavy tailed (as compared with Normal) but that  promote a
qualitative different behavior of the estimators. The three priors have
the same origin given by an Empirical Bayes analysis, and the scales have
been matched by equating the interquartile ranges, to ease the comparison.

\subsection{Full Bayesian Strategy}

\subsubsection{Model 4: Full Bayesian conjugate
model}

This model assigns vague conjugate priors to the common mean $M$ and
the common variance $\sigma^2$. This is the fully Bayes version of
Efron and Morrris Empirical Bayes model.

\begin{eqnarray*}
X_i & \sim & \mbox{Normal}(\mu_i,1), i=1,\ldots,18\\
\mu_i & \sim & \mbox{Normal}(\mbox{M},\sigma^2)\\
M & \sim & N(0,10^5), \;\; \sigma^2  \sim  \mbox{Inv-Gamma}(0.01,0.01)
\end{eqnarray*}

To make Model 4 robust, it is not enough to change the Normal Prior assumption. It is also necessary to replace the assumption about the scales the Inverted-Gamma distribution, see Gelman (2006). We first present the alternative models and delay until next section the discussion of the Scaled-Beta2 model, SBeta2.
\subsubsection{Model 5: Normal likelihood, Double Exponential prior for $\mu_i$, vague Double Exponential prior for the general mean $M$,
Scaled Beta2$(1,1,1)$ prior for the location parameter
$\sigma=\frac{\nu}{\sqrt{2}}$}
\begin{eqnarray*}
X_i & \sim & \mbox{Normal}(\mu_i,1), i=1,\ldots,18\\
\mu_i & \sim & \mbox{DE}(\mbox{M},\sqrt{2}\sigma)\\
M & \sim & \mbox{DE}(0,\sqrt{2}\times 10^{3}), \;\;
\sigma  \sim  \mbox{Beta2}(1,1,1)
\end{eqnarray*}

\subsubsection{Model 6: Normal likelihood, Cauchy prior for $\mu_i$, vague Cauchy prior for the general mean $M$,
Scaled Beta2$(1,1,4)$ prior for the scale parameter}
\begin{eqnarray*}
X_i & \sim & \mbox{Normal}(\mu_i,1), i=1,\ldots,18\\
\mu_i & \sim & \mbox{Cauchy}(\mbox{M},\sigma)\\
M & \sim & \mbox{Cauchy}(0,10^{3}),\;\; \sigma  \sim  \mbox{ScBeta2}(1,1,4)
\end{eqnarray*}

\subsubsection{Model 7: Normal likelihood, Cauchy prior for $\mu_i$, vague Cauchy prior for the general mean $M$,
Scaled Beta2$(1,1,4)$ prior for the squared scale parameter}
\begin{eqnarray*}
X_i & \sim & \mbox{Normal}(\mu_i,1), i=1,\ldots,18\\
\mu_i & \sim & \mbox{Cauchy}(\mbox{M},\sigma)\\
M & \sim & \mbox{Cauchy}(0,10^{3}),\;\; \sigma^2  \sim  \mbox{ScBeta2}(1,1,4)
\end{eqnarray*}
\subsection{The Scaled Beta2 distribution as a prior for scales}
In this three last models, robust priors (Double Exponential and Cauchy)
have been assigned for the locations. On the other hand we propose the use
of the {\bf scaled Beta distribution of the second kind} with parameters $p$, $q$ and
$b$ family ($\mbox{ScBeta2}(p,q,b)$) as priors for the scale
parameters or for squares of scales. Let $Y$ be a random variable  such that $Y \sim
\mbox{Beta2}(p,q,b)$; its density function \citep[see][]{JKB95}, P\'erez and Pericchi (2009), F\'uquene, P\'erez and Pericchi (2011) is given by:

\[ p(y|p,q)=\frac{\Gamma(p+q)}{\Gamma(p) \Gamma(q)}\frac{1}{b}\cdot
\frac{\left(\frac{y}{b}\right)^{p-1}}{\left(1+\frac{y}{b}\right)^{(p+q)}}, y>0, \]

This family has a very natural justification as a prior for variances in
hierarchical models, as it is obtained as a scale mixture of Gamma
distributions, through a Gamma mixing distributions, in much the same way
that the Student-t is obtained as a scale mixture of Normal distributions,
see \citet{PP09}. The usual prior assumed for scales is the Inverted
Gamma family with very large prior variance. This practice
has come under criticism by \citet{Gel06}, who among other alternatives
propose a half-Cauchy prior. The Beta2
prior has flexible tail behavior which makes it particularly
suitable for modeling. When p = q = 1, the ScBeta2(1,1,b) prior
is very close to the half-Cauchy, so we use it here.


What is the assumed prior for the location parameter in Model 6 after integrating out the scale? This is a novel distribution, to  the best of our knowledge, that we call Cauchy-Scaled-Beta 2, which deserves on its own a special note.\\
\begin{defin}[Cauchy-Scaled Beta2 Prior]

Assume that $\theta|\sigma \sim \mbox{Cauchy}(0,\sigma)$, and
$\sigma \sim \mbox{Beta2}(p,q,b)$, then $\theta$ is distributed as a $\mbox{Cauchy - Scaled Beta2}(0,p,q,b)$ .
\end{defin}

It is remarkable that
when $p=q=1$ (assignment that makes the Beta2 to have Cauchy tails) the marginal density for $\theta$
has an explicit formulae (after a long integration by simple fractions):
\begin{res}
The marginal density for the location $\theta$ of a Cauchy-scaled Beta2 prior is:\\
\begin{eqnarray*}
\pi(\theta) & = & \int_0^{\infty} \frac{b \tau }{\pi(b +\tau )^2 \left(\theta^2+\tau^2\right)} \, d\tau \\
& = & \frac{b}{\pi\left(b^2+\theta^2\right)^2} \left[-
\left(b^2+\theta ^2-\pi b |\theta| \right)+(b -\theta )
(b+\theta ) \left( \log(b)-\log(|\theta|)\right)\right]
\end{eqnarray*}
\end{res}

In Figure 4, the new prior (7) is displayed
along with the Cauchy, Double-Exponential and Normal. This prior
enjoys a number of features that explain why it is optimal in
predicting the batters averages. This prior is unbounded as $\theta
\rightarrow 0$ and has tails even heavier than Cauchy. In a recent
paper \citet{CPS10} \citep[see~also][]{PS10} find such
characteristics for a prior to be both robust and leading to
efficient estimation (they propose particular prior which does not
have an explicit form and that they call ``horseshoe" prior).
Furthermore, the Cauchy-scale-Beta2 besides obeying such desiderata, has
an explicit form (at least for p=q=1) which makes it amenable for
mathematical analysis. We do not know of any other explicit
``horseshoe" prior. (In P\'erez and Pericchi (2011) results for other values of hyper-parameters are obtained)\\
The Scaled Beta2 also permits to model the square of the scale, and
closed forms are also available. This is not a Horseshoe prior since
it does not have a pole at the origin.

\[
\pi(\theta|\mu,\tau,b)=\frac{1}{\pi \sqrt{b}\tau}\cdot
[1+\frac{(\theta-\mu)^2}{b \tau^2}]^{-1},
\]
and
\[
\tau^2 \sim SBeta2(\tau^2|1,1,b).
\] What would be the marginal for $\theta$? It is NOT a Horseshoe
prior. Instead: \\
\textbf{Result 4}
 \begin{equation}
 \pi(\theta)=\frac{1}{2 \sqrt{b} \cdot
 (1+\frac{|\theta-\mu|}{\sqrt{b}})^2}.
 \end{equation}
 This is an interesting marginal on itself, close to a Cauchy,
 and it does not have a pole at zero, so it is not a Horseshoe
 prior. We call it a Cauchy-Scale2 Beta2 prior, which is also suitable
 for Robust Bayesian modeling.\\
 In F\'uquene, P\'erez and Pericchi (2011) this prior is studied and applied in detail. In
 fact a general result for the marginal of the location, for any $p$ and $q$ is obtained in terms of
 the Hypergeometric Function.\\
 \textbf{Assessments of Hyper-parameters in the Scaled Beta 2 Distribution:} We assessed $p=q=1$ in model 6 and 7, which is a sensible default assumption since then, both the value of the scale and its reciprocal is a constant at zero, and both tails are very heavy. We also assumed $b=4$, which is larger than three times the estimator of the between variance, and the results for larger values were found to be quite similar to those with $b=3$.


\ssection{MODEL PREDICTIONS}
Table 2 shows the bating averages predicted for each of the models, and in Figure 5 we display the milder shrinkage of the extremes using robust priors, particularly Cauchy priors, when compared to that of Model 1 and Model 4.
Robustifying the priors pays dividends twice: the relative shrinkage
of the extremes is lower and at the same time the error of prediction is diminished up to $7\%$, for Model 6 and 7 which uses the Cauchy-Scaled Beta2 and the Cauchy-Scale2 Beta 2 priors respectively. Model 3 for example,
which incorporates the Cauchy prior has
decreased 5\% the square prediction error than Model 1, and predicts
for Clemente a more respectful 0.314 average, much higher than the
0.290 from Model 1. Something similar may be said for Model 6 and Model 7.
 On the other hand, the price paid seems less than modest:
computationally now there exist approximations and MCMC
algorithms that make the computations routine.
Thus by a very modest cost in computation, the
robustified model has achieved both goals, decreasing the MSE and
solving or at least alleviating the Clemente problem.


In terms of alternative approaches of Statistics that merge direct and indirect evidence, the difference between (sensible and objective versions) of Empirical and Fully Bayes Hierarchical Modeling is relatively small as compared with the difference between heavy and light tail priors.

In general the Clemente problem is closely related to the implicit dogmatism inherent, not in Bayes in general, but in ``Conjugate Bayes with Square Loss". The way out seems to be: either Empirical or Fully Bayesian Hierarchical Model, but making emphasis on Robustness.

\ssection{THE RESURGENCE OF ``OBJECTIVE ROBUST BAYESIAN ANALYSIS"}
It can be argued that there is a resurgence of Objective Robust Bayesian
Analisis (ORBA). To put this article in perspective we should mention some
recent contributions. The first, even though its approach is subjective,
is \citet{And06}, on which the theory of Regularly Varying (RV) functions
is used to check robustness, for general location and scale parameters.
The second is \newline \citet{FCP09} on which it is proved the Generalized
Polynomial Tails Comparison (GPTC) Theorem, and the properties of Berger's
prior are analyzed (prior mentioned in the Polson and Scott's paper). In
\citet{PP09}, the relationship between RV and GPTC is established, and
naturally the Beta of the Second Kind appears in a Meta-analysis of
different hospitals. On the other hand, \citet{CPS10} introduced the
``Horseshoe Estimator" which obeys certain desiderata convenient for
robust analysis in hierarchical models.

In the present article we add the following insights: 1) We established a bridge between robust losses and robust priors.
The theoretical duality between losses and priors has been mentioned before, for example by \citet[p.~161]{Ber85}  but to the best of our
knowledge the consequences of looking at robust procedures with the glass of robust penalties is new. Furthermore note that the way on which the bridge between
 priors and Losses is done, is through an empirical Bayes component, since the loss is centered and scaled through an EB reasoning 2) We introduce here a robust
prior for scales, the Scaled Beta type 2, that is a convenient alternative to the over-used Inverted-Gamma prior. Furthermore we show the novel Cauchy-scale-Beta2 prior which is an explicit objective prior that obeys the desiderata of a ``Horseshoe" prior, as well as the Cauchy-Scaled2 Beta2 prior with general closed form marginal distributions for location parameters. 3) We illustrate, using a classical data set, that it is possible
to alleviate the ``Clemente problem" and at the same time reduce mean square error of prediction as compared with non-robust conjugate approaches and with the James-Stein estimator.
4) We illustrate that the differences between sensible versions of Empirical Bayes and Objective Bayes are relatively small in practice.
 Much more important is the difference between robust Bayes and conjugate Bayes.
 This coupled with the fact that Robust Bayes is in general less dogmatic,  in the sense that in conflict prior information is discarded,
 makes Objective Robust Bayes a strong candidate for statistical synthesis.

 The scope of applications of Robust Objective Procedures is wide ranging, and includes any problem calling for the use of hierarchical models. Just to mention a few, in the analysis of population dynamics for the orchid genus {\em Caladenia} presented in \citet{TP10}, hierarchical models based on vague conjugate priors were used, but even though model selection procedures clearly pointed towards a hierarchical model as the selected one, it gave results that were not biologically sound, probably due to excessive shrinkage towards the species with more data.  Another example is the unfair ``pulling down" of hitherto perfect score hospitals in hospital profiling, just because there are a couple of hospitals with bad performance \citep[see][]{NS07}. \emph{Shrinkage is a good thing, but non-robust methods yields too much of a good thing}.

\newpage
\begin{table}[tbhp]
  \centering
{\small
  \begin{tabular}{lccc}
    \hline
    &Batting average &Batting average &At bats \\
Player& for first 45 &for remainder&for remainder\\
& at bats & of season & of season\\
\hline    \hline
Clemente (Pitts, NL)&0.400&0.346&367\\
F. Robinson (Balt, AL)&0.378&0.298&426\\
F. Howard (Wash,AL)&0.356&0.276&521\\
Johnstone (Cal, AL)&0.333&0.222&275\\
Berry (Chi, AL)&0.311&0.273&418\\
Spencer (Cal, AL)&0.311&0.270&466\\
Kessinger (Chi, NL)&0.289&0.263&586\\
Alvarado (Bos, AL)&0.267&0.210&138\\
Santo (Chi, NL)&0.244&0.269&510\\
Swoboda (NY, NL)&0.244&0.230&200\\
Unser (Wash, AL)&0.222&0.264&277\\
Williams (Chi, AL)&0.222&0.256&270\\
Scott (Bos, AL)&0.222&0.303&435\\
Petrocelli (Bos, AL)&0.222&0.264&538\\
E. Rodriguez (KC, AL)&0.222&0.226&186\\
Campaneris (Oak, AL)&0.200&0.285&558\\
Munson (NY, AL)&0.178&0.316&408\\
Alvis (Mil, NL)&0.156&0.200&70\\
    \hline
  \end{tabular}
  }
  \caption{Original data: $1970$ batting averages for $18$ MBL players }\label{tab:data}
\end{table}

\clearpage

\begin{sidewaystable}[tbhp]
\begin{center}
{\footnotesize \begin{tabular}{lcccccccccc} \hline
Player&Observed &First 45 & General &Model 1 &Model 2& Model3& Model 4 & Model 5& Model 6 & Model 7\\
&season& (MLE)& mean &&  &&& & &\\

\hline
Clemente&0.346&0.400&0.265&0.290&0.304&0.314&0.282&0.298&0.291&0.309\\
Robinson&0.298&0.378&0.265&0.286&0.296&0.301&0.279&0.291&0.283&0.296\\
Howard&0.276&0.356&0.265&0.282&0.288&0.291&0.277&0.285&0.2762&0.287\\
Johnstone&0.222&0.333&0.265&0.277&0.281&0.282&0.273&0.279&0.272&0.279\\
Berry&0.273&0.311&0.265&0.273&0.275&0.275&0.270&0.273&0.269&0.273\\
Spencer&0.270&0.311&0.265&0.273&0.275&0.275&0.270&0.273&0.269&0.273\\
Kessinger&0.263&0.289&0.265&0.269&0.269&0.270&0.267&0.268&0.266&0.269\\
Alvarado&0.210&0.267&0.265&0.265&0.264&0.264&0.264&0.264&0.263&0.263\\
Santo&0.269&0.244&0.265&0.259&0.258&0.259&0.261&0.259&0.260&0.259\\
Swoboda&0.230&0.244&0.265&0.259&0.258&0.259&0.260&0.259&0.260&0.259\\
Unser&0.264&0.222&0.265&0.255&0.252&0.254&0.257&0.254&0.258&0.253\\
Williams&0.256&0.222&0.265&0.255&0.252&0.254&0.257&0.254&0.257&0.254\\
Scott&0.303&0.222&0.265&0.255&0.252&0.253&0.257&0.254&0.257&0.253\\
Petrocelli&0.264&0.222&0.265&0.255&0.252&0.253&0.257&0.254&0.257&0.253\\
Rodriguez&0.226&0.222&0.265&0.255&0.252&0.253&0.257&0.254&0.257&0.254\\
Campaneris&0.285&0.200&0.265&0.250&0.245&0.247&0.254&0.248&0.254&0.247\\
Munson&0.316&0.178&0.265&0.245&0.237&0.238&0.251&0.242&0.249&0.240\\
Alvis&0.200&0.156&0.265&0.240&0.228&0.226&0.247&0.234&0.242&0.230\\
\hline
{\bf MSE ($x10^3$)}&&{\bf 4.184}& {\bf 1.348} & {\bf 1.196} & {\bf 1.187} & {\bf 1.137} & {\bf 1.198} &{\bf 1.168}
& {\bf 1.108}& {\bf 1.117}\\
$\frac{MSE(\mbox{Model})}{MSE(\mbox{Model 1})}$&& {\bf 350\%} &{\bf 113\%} & {\bf 100\%} & {\bf 99\%} & {\bf 95\%} & {\bf 100\%} & {\bf 98\%} & {\bf 93\%} & {\bf 93\%}\\
\end{tabular} }
\caption{Estimators and mean square error of prediction for MLE, general
mean, empirical Bayes models and full Bayesian models}
\end{center}
\end{sidewaystable}

\clearpage

\noindent
Figure 1: Quadratic loss and Cauchy over Gaussian loss ($\delta_i=M$).
\vspace{0.2cm}

\noindent
Figure 2: Effect of the conflict between the observed value $x_i$ and the prior location $M$ on the difference between the posterior mean and the observed value ($x_i-E(\mu|x)$). Normal linear unbounded influence of prior location, monotone limited translation in Double Exponential and discarding influence in the Cauchy can be observed in this figure.
\vspace{0.2cm}

\noindent
Figure 3:  Posterior expectations for Normal, Double Exponential and Caucy priors, Efron and Morris' baseball data set. Transformed observations are displayed on the X-axis and prior location is fixed at the general average ($M=\bar{X}$). Different shrinkage behavior can be observed for Normal, Double Exponential and Cauchy priors
\vspace{0.2cm}

\noindent
Figure 4: Comparison of Cauchy-Scaled Beta2, Normal, Double Exponential and Cauchy distributions, at the center (upper figure) and at the tail (lower figure).
\vspace{0.2cm}

\noindent
Figure 5: Comparison of shrinkages: upper figure shows observed data, Efron and Morris estimates and Robust Empirical Bayes estimates. Lower figure shows observed data, non-robust Full Bayes estimates and Robust Full Bayes estimates.

\begin{figure}[p]
  \begin{center}
  \includegraphics[scale=0.8]{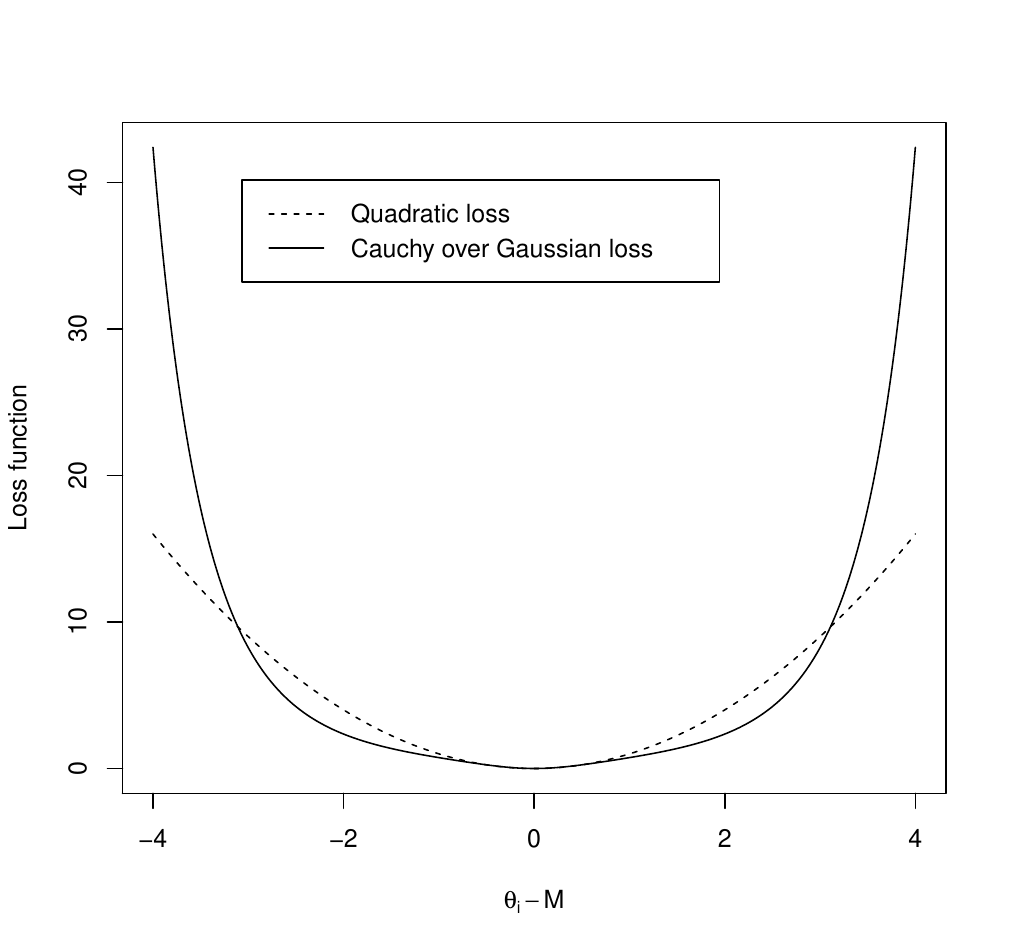}
  \end{center}
\end{figure}

\begin{figure}[p]
  \begin{center}
  \includegraphics{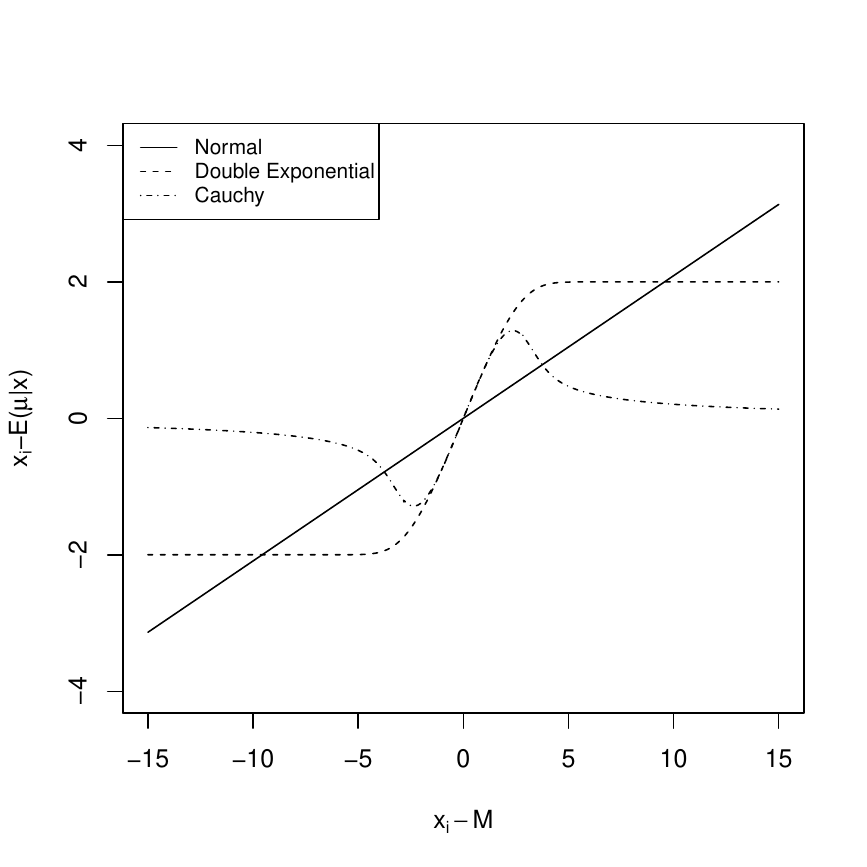}
  \end{center}

\end{figure}

\begin{figure}[p]
  \begin{center}
  \includegraphics{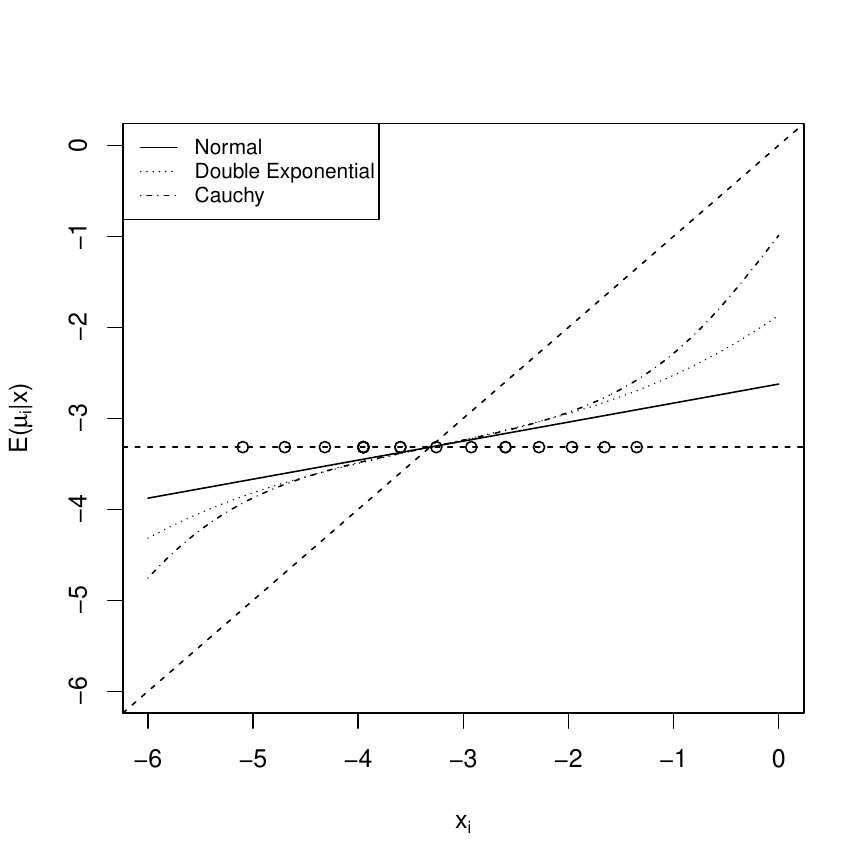}\\
  \end{center}
\end{figure}

\begin{figure}[p]
\begin{center}
\begin{tabular}{c}
\begin{minipage}{5in}
\centering
\includegraphics[scale=0.70]{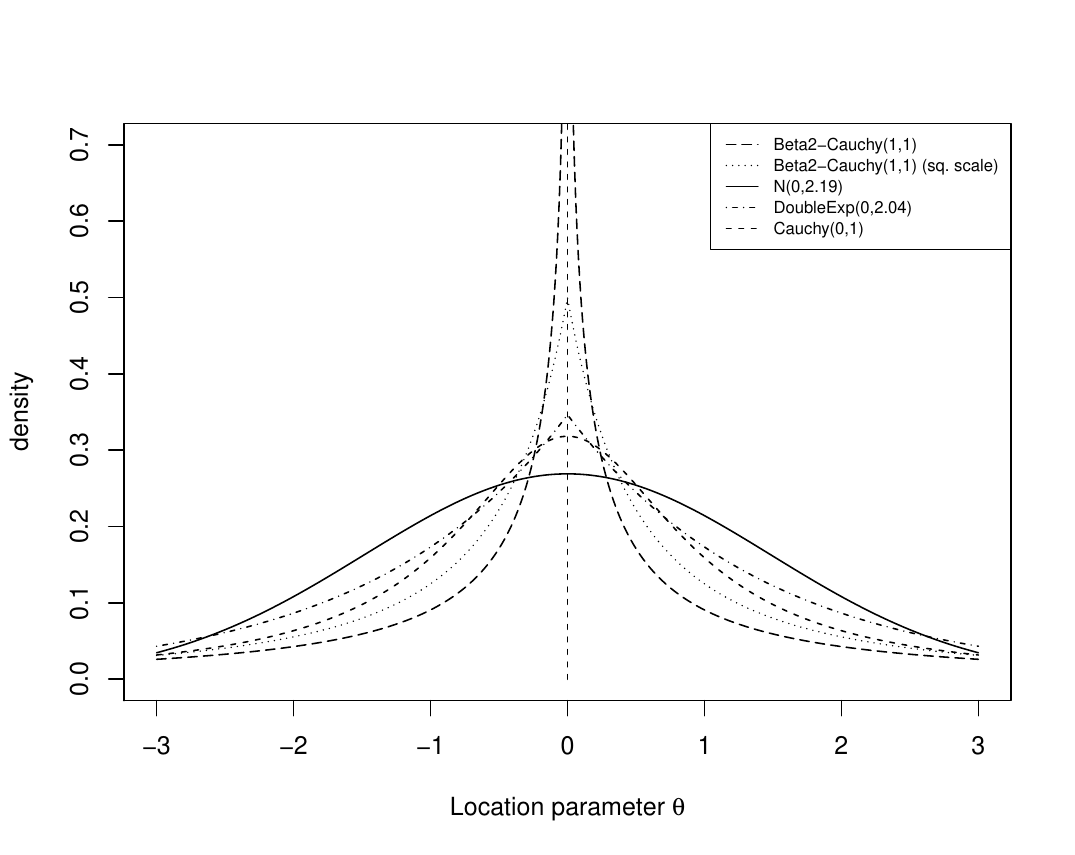}
\end{minipage}
\\
\begin{minipage}{5in}
\centering
\includegraphics[scale=0.70]{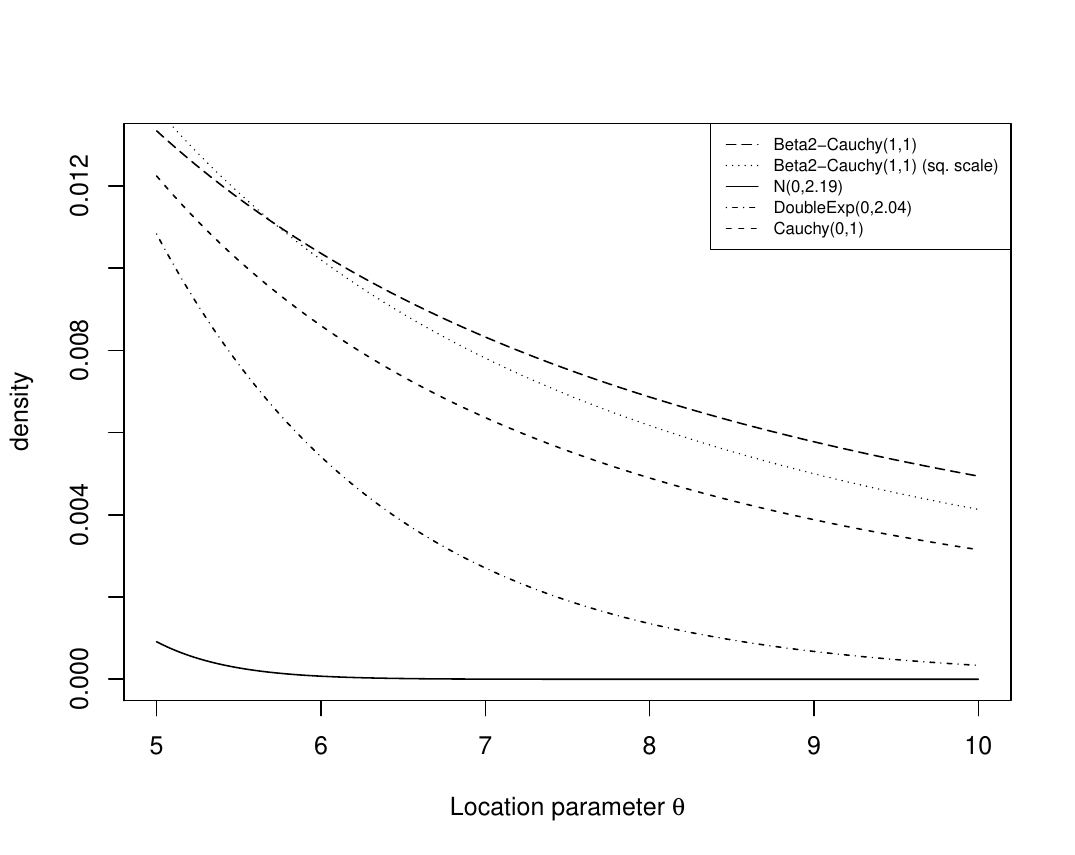}
\end{minipage}

\end{tabular}
\end{center}
\end{figure}

\begin{figure}
\begin{center}
\begin{tabular}{c}
\begin{minipage}{5in}
\centering
\includegraphics[scale=0.65]{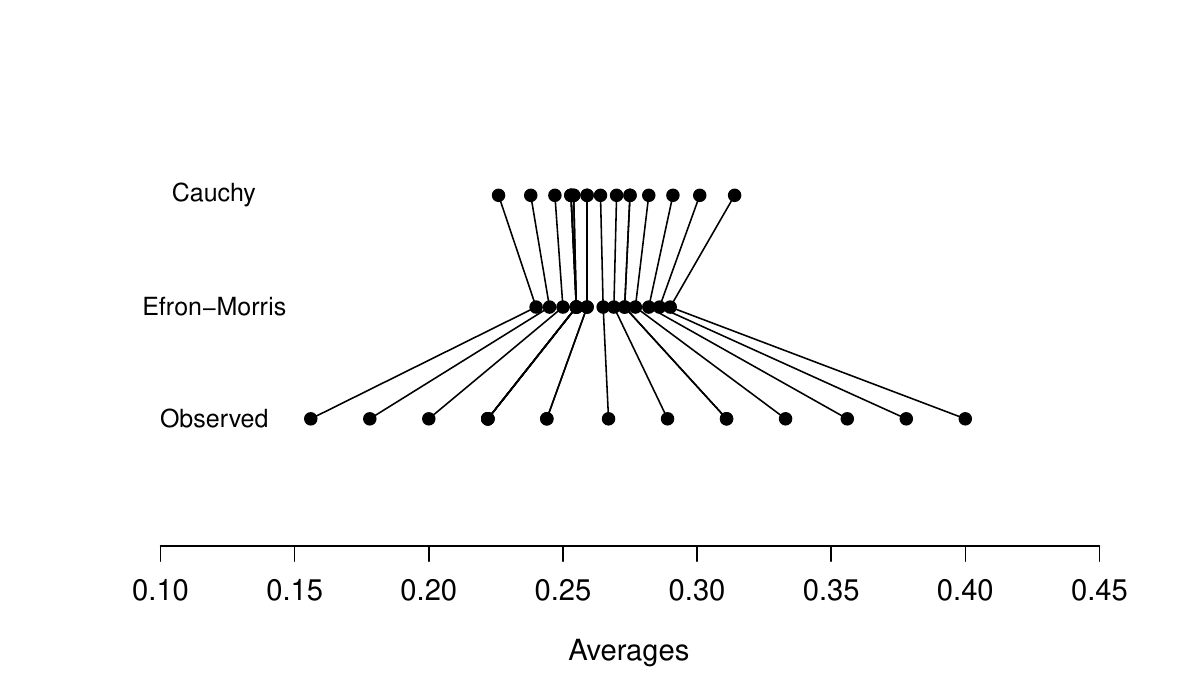}
\end{minipage}
\\

(a)MLE, Model 1, and Robust Empirical Bayes Model 3.\\

\begin{minipage}{5in}
\centering
\includegraphics[scale=0.65]{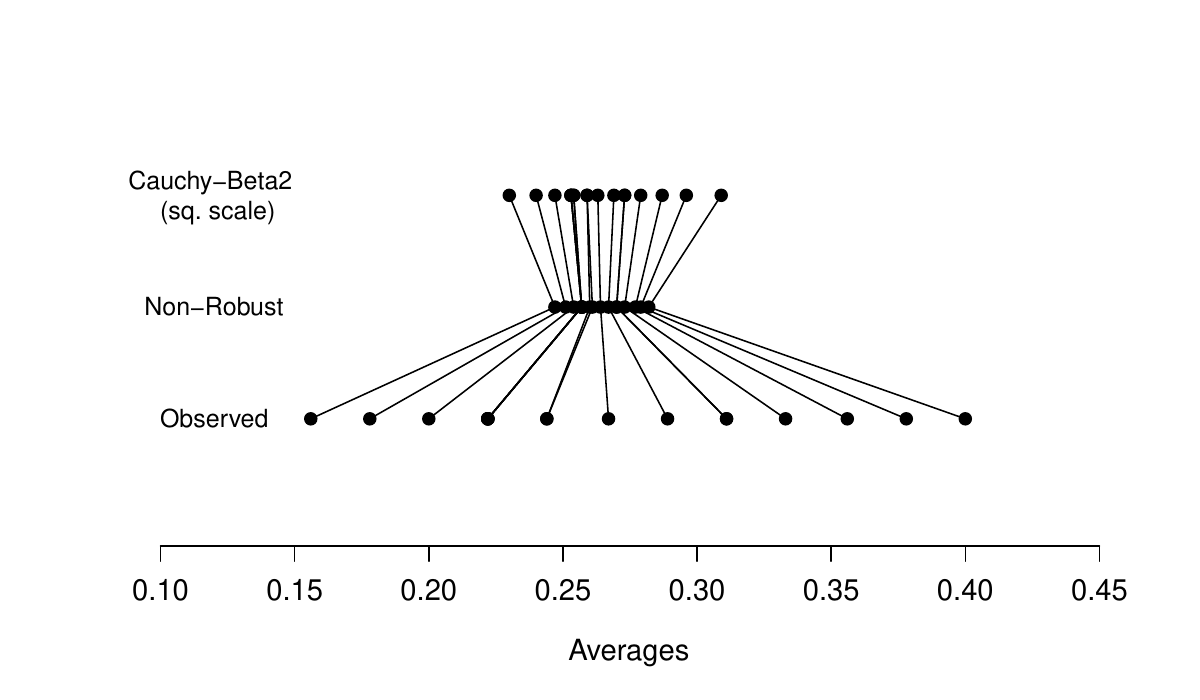}
\end{minipage}
\\
(b)MLE, Model 4, and Robust Full Bayes Hierarchical Model 7.

\end{tabular}
\end{center}
\end{figure}

\end{document}